\documentclass[twocolumn,superscriptaddress,showpacs,aps,prb]{revtex4}
\usepackage{makeidx}
\usepackage{bm}
\usepackage{graphics}
\usepackage[dvips]{epsfig}
\newcounter{fig}

\begin{document}
\title{Features of interband absorption in narrow-gap semiconductors}
\author{L.A. Falkovsky}
\affiliation{L.D. Landau Institute for Theoretical Physics, Moscow
117334, Russia} \affiliation{Institute of the High Pressure
Physics, Troitsk 142190, Russia}
\pacs{71.20.Nr, 78.20.Ci, 78.20.Bh}

\date{\today}      

\begin{abstract}
For  semiconductors and semimetals possessing a narrow gap between
bands with different parity, the dispersion of the dielectric
function is explicitly evaluated in the infrared region. The
imaginary part of the dielectric function has a plateau above the
absorption threshold for the interband electron transitions. The
real part of the dielectric function has a logarithmic singularity
at the threshold. This results in the large contribution into the
dielectric constant for pure semiconductors at low frequencies.
For samples with degenerate carriers, the real part of the
dielectric function is divergent at the absorption threshold. This
divergence is smeared with the temperature or the collision rate.
\end{abstract}
\maketitle

Usually experiments and theories describe \cite{KYP,APP,XWS,AFZ}
the direct  allowed  transitions in terms of the Fermi golden rule
which provides  the imaginary part of the dielectric function (or
the real conductivity)
\begin{equation}\label{wn}
\epsilon^{\prime\prime}(\omega)\sim\int|d_{vc}|^{2}
\delta[\varepsilon_{c}(\mathbf{p})-
\varepsilon_{v}(\mathbf{p})-\hbar\omega]\frac{2d^{3}p}{(2\pi)^{3}}
 ,\end{equation} giving the square
root dependence \(\epsilon^{\prime
\prime}(\omega)\sim\sqrt{\hbar\omega-2\varepsilon_{g}}\)\, near
the band edge absorption for the case when the conduction band  is
empty and the valence band  is filled. The electron-hole Coulomb
interaction smears this square-root singularity. For  doped
semiconductors, the threshold of absorption  is determined by the
carrier concentration, i.e., the chemical potential $\mu$ if the
temperature is low enough.  The interband transitions of carriers
give also a contribution into the real part
$\epsilon^{\prime}(\omega)$ which can be calculated with the help
of the Kramers--Kronig relations. In reality, these numerical
calculations involve the pseudopotential form-factor and do not
present an evident result (see, for instance, Ref.
\cite{KYP,APP}). It is more productive to use an explicit
expression for the complex optical conductivity which can be
derived using the Kubo formula or the RPA approach.

Here, we present  calculations of the dielectric function for an
important case when the gap $\varepsilon_g$ between the conduction
and valence bands is much smaller than the distance $\varepsilon
_{at}$ (on the atomic scale) to  other bands. The model is
applicable to the  IV-VI  semiconductors (as PbTe, PbSe, and PbS),
i.e., such narrow-gap semiconductors and semimetals, where the
narrow gap appears as a result of  intersections of two bands with
different parity. We evaluate the dispersion of the real part of
the dielectric function and the reflectance along with the
behavior of the imaginary part around  the absorption threshold.
We find that the contribution of the electron transitions into the
real part has the logarithmic singularity at the threshold and can
be more essential for optical properties than absorption given by
the imaginary part of the dielectric function.

The effective Hamiltonian of the problem can be written as a
4$\times$4 matrix \cite{Da}:
\begin{equation}\begin{array}{cccc}
 H =&\left(
\begin{array}{cc}
            \varepsilon_{g} & H_{1} \\
          H_{1}^{+} & - \varepsilon_{g}
           \end{array}\right)\,, \quad
&H_{1} = &\left( \begin{array}{cc}
           v_{l} p_{z} & v_{t}p_{-} \\
           v_{t}p_{+} & -v_{l} p_{z}
           \end{array} \right)\,,\end{array}
\label{ham}\end{equation} where $p_{\pm}=p_x\pm ip_y$ and $v_t,
v_{\ell}$ are constants. For the bands of different parity, the
terms linear in the quasi-momentum $p_j$ appear only in the
off-diagonal matrix elements. The quadratic terms can  be added to
$\varepsilon_g$ on the main diagonal. We omit these terms because
their contribution has the order of
$\varepsilon_g/\varepsilon_{at}\ll 1$.

The Hamiltonian has the two-fold (due to spin) eigenvalue
\(\varepsilon_1({\bf
p})=[\varepsilon_g+v^2_{\ell}p_z^2+v_t^2p_{\perp}^2]^{1/2}\)
and the two-fold eigenvalue \(\varepsilon_2=-\varepsilon_1({\bf
p})\).

We use the general expression for the conductivity obtained in
Ref.\cite{FV}, where  two-dimensional graphene was considered.
That expression is also applicable in the three-dimensional case.
For the optical range, when the frequency is large in comparison
with both the spacial dispersion of light ($\omega\gg kv$) and the
collision rate of carriers $\nu$, the complex conductivity has the
form
\begin{eqnarray}\label{con}
\sigma _{ij}(\omega ) = \frac{-ie^{2}}{4\pi ^{3}\omega} \left\{
\sum_{a=1,2}\int \frac {d f(\varepsilon_a)}{d\varepsilon}
v^{i}_{11}v^{j}_{11}d^{3}p\right.   \\
\left. -2\omega^2 \int \frac{
[f(-\varepsilon_{1})-f(\varepsilon _{1})]}{%
\varepsilon _{1}(\omega ^{2}-4\varepsilon_{1}^2)}
(v_{13}^{i}v_{31}^{j}+v_{14}^{i}v_{41}^{j})d^{3}p\right\}\,,
\nonumber
\end{eqnarray}%
where  ${\bf v}=\partial H/\partial {\bf p}$ is the velocity
operator and $f(\varepsilon)$ is the Fermi function. Here, the
first term is the known Drude-Boltzmann intraband conductivity. If
the collision rate $\nu$ of carriers is taken into account, we
have to substitute $\omega\rightarrow\omega+i\nu$. The second
integral is given by the interband electron transitions.  The
subscripts 3 and 4 correspond with two states of the Hamiltonian
in the valence band, whereas the subscript 1 corresponds with the
given state in the conduction band.

The matrix elements of velocity $v_{nm}^{j}$ should be calculated
in the representation, where the Hamiltonian (\ref{ham}) has a
diagonal form.  The operator  transforming the Hamiltonian to this
form can be written as following
\[{ U} = \left( \begin{array}{cccc}
 k_{z}/n_1 & k_{-}/n_1 &k_{z}/n_2 & k_{-}/n_2 \\
 k_{+}/n_1& -k_{z}/n_1  &k_{+}/n_2 &- k_{z}/n_2 \\
 a_1& 0 &-a_2& 0 \\
 0 & a_1& 0 &-a_2
 \end{array} \right)\,, \]
where \(k_z=v_{\ell}p_z\, , k_{\pm}=v_t p_{\pm}\)\,,
\(n_{1,2}=\sqrt{2\varepsilon_1(\varepsilon_1\mp\varepsilon_g)}\)\,,
\(a_{1,2}=\sqrt{(\varepsilon_1\mp\varepsilon_g)/2\varepsilon_1}\).
In this representation, the velocity operator has the following
matrix form
\[U^{-1}\mathbf{v}U =\left( \begin{array}{cccc}
            \mathbf{v}_{11} & 0 & \mathbf{v}_{13} & \mathbf{v}_{14}\\
            0 & \mathbf{v}_{11} & -\mathbf{v}_{14}^{*} & \mathbf{v}_{13}^{*}\\
            \mathbf{v}_{13}^{*} & -\mathbf{v}_{14} & -\mathbf{v}_{11} & 0\\
            \mathbf{v}_{14}^{*} & \mathbf{v}_{13} & 0 & -\mathbf{v}_{11}
           \end{array} \right)\,, \]
where
\[\begin{array}{c}
\mathbf{v}_{11}=\partial \varepsilon_1/\partial {\bf p}\,,\\
\mathbf{v}_{13}=-2\{\varepsilon_g[v_{\ell}^2p_z{\bf
e}_z+v_t^2(p_x{\bf e}_x+p_y{\bf
e}_y)]\\+i\varepsilon_gv_t^2(p_x{\bf e}_y-p_y{\bf
e}_x)\}/n_1n_2\,,\\
 \mathbf{v}_{14}=2v_{\ell}v_t\varepsilon_1(p_{-}{\bf e}_z-p_{z}{\bf
e}_{-})/n_1n_2\,,
\end{array}\]
and ${\bf e}_j$ are the unit vectors directed along the
corresponding coordinate axes.

 Calculations give the velocity
squared $(v^{x}_{11})^2=v_t^4p_x^2/\varepsilon_1^2$ for the
intraband conductivity and the sum
\[v_{13}^{x}v_{31}^{x}+v_{14}^{x}v_{41}^{x}=
v_t^2(1-v_t^2p_x^2/\varepsilon_1^2)\]
which presents indeed  the dipole matrix elements squared for the
interband electron transitions. Introducing the variables
  ($\varepsilon,\theta, \varphi$) of integration,
$p_z=\sqrt{\varepsilon^2-\varepsilon_g^2}\cos{\theta}/v_{\ell}\,,
p_{\perp}=\sqrt{\varepsilon^2-\varepsilon_g^2}\sin{\theta}/v_{t}$,
and integrating over the angles, we find that only diagonal
elements $\sigma_{jj}$ of the conductivity tensor are not
vanishing for this symmetrical model. Then, we can write the
intra- and inter-band conductivities in the following form
\begin{eqnarray}\label{intra}
\sigma _{xx}^{intra}(\omega ) = \frac{-ie^{2}}{3\pi
^{2}v_{\ell}\omega}\\
\times\int_{\varepsilon_g}^{\varepsilon_{at}}[f^{\prime}(\varepsilon)+
f^{\prime}(-\varepsilon)]
(\varepsilon^2-\varepsilon_g^2)^{3/2}\frac{d\varepsilon}{\varepsilon}\,,
\nonumber
\end{eqnarray}

\begin{eqnarray}\label{inter}
 \sigma _{xx}^{inter}(\omega ) = \frac{2ie^{2}\omega}{3\pi
 ^{2}v_{\ell}}\\ \times
 \int_{\varepsilon_g}^{\varepsilon_{at}} [f(-\varepsilon)-f(\varepsilon )]\frac{
(\varepsilon^2-\varepsilon_g^2)^{1/2}}{%
\omega ^{2}-4\varepsilon^2}
\left(1+\frac{\varepsilon_g^2}{2\varepsilon^2}\right)d\varepsilon\,.
\nonumber
\end{eqnarray}%
The real part of the last integral diverges logarithmically at the
upper limit, where our linear expansion of the Hamiltonian
(\ref{ham}) does not work. But the main contribution into the
integral comes from the region $\varepsilon_g\ll\varepsilon\ll
\varepsilon_{at}$. Therefore, we can cut off the integral with the
atomic parameter $\varepsilon_{at}$ of the order of several eV.

In the limiting case  $(T, \varepsilon_g, \nu)=0$, the integrals
(\ref{intra}) and (\ref{inter})  give a very simple result:
\begin{eqnarray}\label{ans}
\sigma_{xx}(\omega)=\frac{e^{2}}{3\pi^{2}\hbar v_{\ell}}\\
\nonumber\times \left[i\frac{\mu^2}{\omega}
-\frac{i}{4}\omega\ln{\frac{4\varepsilon_{at}^2}{|\omega^2-4\mu^2|}}+
\frac{\pi}{4}\omega\,\theta(\omega-2\mu)\right]\,,
\end{eqnarray}
where the chemical potential $\mu$ is scaled from the middle  of
the gap. The Drude-Boltzmann conductivity can be obtained from the
first term in Eq. (\ref{ans}) if we substitute $\omega\rightarrow
\omega+i\nu$. The third term with the step $\theta$-function
presents the interband absorption in the case of the degenerate
carrier statistics. The second logarithmically divergent term is
also a result of the interband electron transitions.

In order to get the total conductivity, we have to summarize the
contributions of all the valleys. For instance, there are four
valleys each around the L points of the Brillouin zone in the
IV-VI semiconductors.   The  component $\sigma_{zz}(\omega)$ of
conductivity  taken in the coordinate frame of the valley is given
by Eqs. (\ref{intra}) and (\ref{inter}) with the substitution
$v_t^2/v_{\ell}$ instead of $v_{\ell}$. Summarizing over the L
points, we find the total conductivity tensor having the diagonal
form, where all the diagonal components $\sigma(\omega)$ are given
by Eqs. (\ref{intra}),  (\ref{inter}), and (\ref{ans}), but the
substitution
\[ \frac{1}{v_{\ell}}\rightarrow
\frac{8}{3v_{\ell}}+\frac{4v_{\ell}}{3v_t^2}\equiv\frac{1}{v}\]
has to be made. Then, the dielectric function can be written in
the usual way:
\begin{equation}
\epsilon(\omega)=\epsilon_0+4\pi i\sigma(\omega)/\omega\,.
\label{df1}
\end{equation}
Because we are interested in low frequencies when the contribution
of the narrow bands into the dielectric constant is leading, we
can put $\epsilon_0=1$.

 The divergence of
the second term in Eq. (\ref{ans}) at the threshold $\omega=2\mu$
is cut off with temperature. Calculations show that  we should
substitute
\[\omega^2-4\mu^2\rightarrow|\omega^2-4\mu^2|+4\omega T\]
 at low, but finite temperatures.
\begin{figure}[]
\noindent\centering{
\includegraphics[width=90mm]{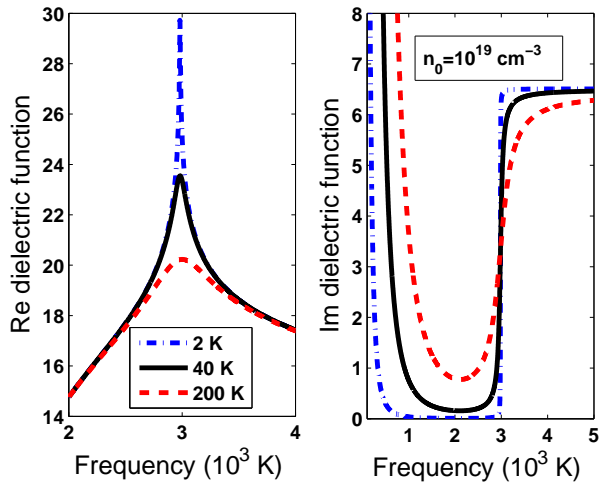}
} \caption{ Real and imaginary parts of the dielectric function
for the scattering rates (in K) listed on the left panel;
 the carrier concentration  $10^{19}$ cm$^{-3}$ corresponds
to the chemical potential $\mu=1491$ K supposed to be much larger
than the gap $\varepsilon_g$; the carrier collision rate takes
values 2 K (dash-dotted line), 40 K (solid line), and 200 K
(dashed line). }\label{ds}
\end{figure}
If the collision rate is more essential than temperature
($\nu>T$), the dispersion of   conductivity  and  dielectric
function is given by Eqs. (\ref{ans}) and (\ref{df1}), where the
substitutions
\begin{eqnarray}\nonumber
\omega^2-4\mu^2\rightarrow[(\omega^2-4\mu^2)^2+(2\omega
\nu)^2]^{1/2}\,,\\ \nonumber \theta(\omega-2\mu)\rightarrow
\frac{1}{2}+\frac{1}{\pi}\arctan{[(\omega-2\mu)/\nu]}
\end{eqnarray}
 should be made.
\begin{figure}[]
\noindent\centering{
\includegraphics[width=90mm]{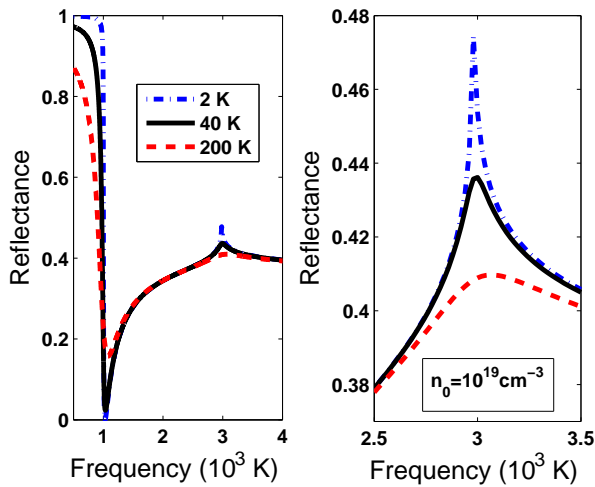}
} \caption{ Reflection coefficient for normal incidence; the
values of parameters are the same as in Fig. 1. }\label{ds}
\end{figure}

The real and imaginary parts of the dielectric function are
plotted in Fig. 1 versus $\omega$ for three values of the
collision rate listed in the left panel and the carrier
concentration 10$^{19}$cm$^{-3}$ which corresponds to the chemical
potential $\mu=1491 $ K supposed to be much larger than
$\varepsilon_g$.   The electron velocities $v_{\ell}=3.8\times
10^7$ cm/s and $v_{t}=5.1\times 10^7$ cm/s are known from
literature for PbSe, PbTe, and PbS. At low frequencies,
$\omega\ll\mu$, the behavior of the dielectric function is
determined by the intraband conductivity. In particular, the
imaginary part of the dielectric function grows with the
increasing collision rate.

Afterwards, the real part of the dielectric function  goes through
zero at the frequency $\omega_0$ determined roughly by the
equation
\begin{equation}
\epsilon_0- \frac{2e^2}{3\pi\hbar
v}\left(\frac{2\mu^2}{\omega^2_0}-\ln\frac{\varepsilon_{at}}{\mu}\right)=0\,.
\label{zeq}\end{equation}

The threshold of absorption (see the right panel) occurs at
$\omega=2\mu$ for the finite carrier concentration. It is smeared
while the collision rate (or temperature) is rising. At this
frequency, the real part of the dielectric function (left panel)
has a peak if the collision rate is low enough. This is an effect
of the interband electron transitions.

Finally, we have to consider the case of the empty conduction band
when the gap $\varepsilon_g$ plays a role. Let temperature $T=0$.
The imaginary part, Eq. (\ref{inter}), is given by
\[ \epsilon^{\prime \prime }(\omega)= \frac{e^{2}}{3\hbar
v\omega^3}(\omega^2+2\varepsilon_g^2)\sqrt{\omega^2-4\varepsilon_g^2}\]
 which agrees with the known result, Eq. (\ref{wn}), near the band
 edge,
 $\omega\rightarrow2\varepsilon_g$, and
 goes to the constant value $\epsilon_0^{\prime\prime}$
 at $\omega\gg\varepsilon_g$. We can
 esimate this constant value using the  parameters $v_l$
 and $v_t$ listed above.
 We obtain $\epsilon_0^{\prime\prime}=6.5$ for all that material
 in excellent agreement with experimental data \cite{KB, SSA}
 interpolated from frequencies
 higher than  0.5 eV. To the best of our knowledge, no experimental data
 exist for frequencies lower than 0.5 eV. Notice that no
 adjustable parameters are used.

The real part of dielectric function  given by the interband
transitions in the case $T\ll(\varepsilon_g,\mu)$ is equal
\[\epsilon^{\prime}(\omega)=\epsilon_0+\frac{2e^{2}}{3\pi\hbar
v} \ln{\frac{\varepsilon_{at}}{max\{ \varepsilon_g, \omega/2\}}}\]
 within  logarithmic accuracy. Emphasize, that in contrast with
 the case of degenerate carriers, Eq. (\ref{ans}), the real part of the
 dielectric function has  the large, but
 convergent value at the band edge $\omega=2\varepsilon_g$.
 Thus, the contribution into the real part $\epsilon_0^{\prime}$
 depends on the
 gap  taking the various values in the IV-VI semiconductors.
 Comparing the values with the constant $\epsilon_0^{\prime\prime}$, we find
 \[\epsilon^{\prime}(\omega)\simeq\frac{2}{\pi}\epsilon^{\prime \prime }_0
 \ln{\frac{\varepsilon_{at}}{max\{ \varepsilon_g, \omega/2\}}}\,.\]
 We see that the real part of the dielectric function decreases
 with frequencies  in the domain $\omega>2\varepsilon_g$,
 where the imaginary part has a plateau.
 Taking the typical size of the gap $\varepsilon_g\simeq 0.1$ eV and a reasonable value
 of the cutoff parameter $\varepsilon_{at}\simeq 8$ eV into account,
 we   estimate  the maximum value
 $\epsilon^{\prime}_{max}\simeq  25,$
 which agrees with the numerical calculations \cite{APP}.

The reflection coefficient
\[R=\left|\frac{\epsilon\cos\theta-\sqrt{\epsilon-\sin^2\theta}}
{\epsilon\cos\theta+\sqrt{\epsilon-\sin^2\theta}}\right|^2\]
is shown in Fig. 2 for  normal incidence, $\theta=0$. At low
frequencies, the reflection demonstrates the metallic behavior for
a sample with carriers. While the frequency increases, reflectance
drops to zero at the frequency $\omega_0$ approximately determined
by Eq. (\ref{zeq}), where  large logarithm due to the interband
transitions becomes  important. Afterwards, the reflectance is
mainly determined by  the interband transitions.  At the
threshold, $\omega=2\mu=2982$~K, corresponding to  the carrier
concentration $10^{19}$ cm$^{-3}$, the sharp peak of reflectance
should be observed for low temperatures (T$\sim$10 K) and low
collision rates (i.e., if the mean free time $\tau > 10^{-13}$ s
for the carrier concentration on the order of 10$^{18}$--10$^{19}$
cm$^{-3}$) due to singular logarithm in the interband
conductivity. Observation of the peak presents a characterization
of  carrier concentration in pure samples. The peak is followed by
decreasing  reflection because the interband absorbtion grows
above the threshold.

{\it In conclusions,} we find  that the real part of the
dielectric function $\epsilon^{\prime}(\omega)$ contains a
singular contribution from the interband electron transitions. At
low frequencies, it gives the large logarithmic term into the
dielectric constant. While increasing the frequency, we obtain the
dispersion of the dielectric function. Near the threshold of the
interband absorption (at $\omega\simeq 2\mu$ for  degenerate
statistics of carriers),  a peak appears  at low temperatures if
the mean free time of carriers is large enough.

This work was supported by the Russian Foundation for Basic
Research (grant No. 07-02-00571).

\end{document}